\documentclass{cs19proc}

\usepackage{amsfonts}
\usepackage{amssymb}

\usepackage{color}
\newcommand{\clem}[1]{\textcolor{black}{#1}}

\editors{G.~A. Feiden}
\publisher{Zenodo}
\conference{The 19th Cambridge Workshop on Cool Stars, Stellar Systems, and the Sun}
\conferencedate{2016}

\title{Tidal dissipation by inertial waves in differentially rotating convective envelopes of low-mass stars}
\author{Mathieu Guenel,$^{1}$ St\'ephane Mathis,$^{1}$ Cl\'ement Baruteau,$^{2}$ Michel Rieutord$^{2}$}

\affiliation{$^{1}$ Laboratoire AIM Paris-Saclay, CEA/DRF/IRFU/SAp - Universit\'e Paris Diderot - CNRS, 91191 Gif-sur-Yvette, France \\ $^{2}$ IRAP, Observatoire Midi-Pyr\'en\'ees, Universit\'e de Toulouse, 14 avenue Edouard Belin, 31400 Toulouse, France}

\shortauthors{Mathieu Guenel, Cl\'ement Baruteau, St\'ephane Mathis \& Michel Rieutord}

\abs{Tidal interactions in close star-planet or binary star systems may excite inertial waves {(their restoring force is the Coriolis force)} in the convective region of \clem{the} stars. The dissipation of these waves plays a prominent role in the long-term orbital and rotational evolution of the bodies involved. If the primary star \clem{rotates} as a solid body, inertial waves have a Doppler-shifted frequency restricted to the range $[-2\Omega, 2\Omega]$ ($\Omega$ being the angular velocity of the star), and they can propagate in the entire convective region. However, turbulent convection can sustain differential rotation with an equatorial acceleration (as in the Sun) or deceleration that modifies \clem{the frequency range and propagation domain of inertial waves and allows corotation resonances for} non-axisymmetric oscillations. {In this work, }we perform numerical simulations of tidally excited inertial waves \clem{in a differentially rotating convective envelope with a conical (or latitudinal) rotation profile.} The tidal forcing \clem{that} we adopt contains spherical harmonics that correspond to the case of a circular and coplanar orbit. We study the \clem{viscous dissipation of the waves} as a function of tidal frequency for various stellar masses and differential rotation parameters, as well as its dependence on the turbulent viscosity coefficient. We compare our results with previous studies assuming solid-body rotation and point out the \clem{potential} key role of corotation resonances in the dynamical evolution of close-in star-planet or binary systems.}

\begin{document}

\maketitle

\section{Introduction}
The tidal potential exerted by a \clem{close-in} planet on a \clem{low-mass} star may excite inertial waves in the \clem{star's} convective envelope \citep[see, e.g.,][]{OgilvieLin2007,BO2009,Mathis2015}. \clem{The energy and angular momentum deposited by these waves drives the ultimate evolution of the orbit and spin of the system} \citep[e.g.,][]{Hut1980, Meibom2005}. {Moreover, the dissipation of tidal inertial waves presents strong variations with the internal structure and rotation of stars (and planets) and with forcing frequencies. These variations have a profound impact on the evolution of the spin and orbit of close-in systems \citep[][]{ADLPM2014,BM2016}}. Convection \clem{may lead to} conical --- solar/antisolar-type --- or sometimes cylindrical differential rotation profiles (\citealp{Schouetal1998,Barnesetal2005,Matt2011, Gastine2014, Brun2015, Varelaetal2016}). Differential rotation strongly modifies the \clem{propagation and dissipation properties} of inertial waves compared to solid-body rotation, as shown \clem{by} \cite{BR2013} for shellular and cylindrical rotation profiles. \clem{In this communication, we expose our first results on tidally excited inertial waves in a differentially rotating convective envelope, focusing on the case of conical rotation. After reviewing the general properties of inertial eigenmodes with conical differential rotation in Sect.~\ref{sec:eigenmodes}, we present in Sect.~\ref{sec:forced} some preliminary results on the tidal dissipation of forced inertial modes.}

\section{Properties of inertial eigenmodes}
\label{sec:eigenmodes}
\clem{We model the convective envelope of low-mass stars as an homogeneous incompressible fluid of density $\rho$ and kinematic {(turbulent)} viscosity $\nu$} inside a spherical shell of aspect ratio $\eta$. \clem{We take the rotation profile of the envelope to be} conical, i.e. depending only on the colatitude $\theta$:
\begin{equation}
\Omega(\theta) = \Omega_{\rm ref}\left(1 + \varepsilon \sin^2 \theta\right),
\label{eq:rotation_profile}
\end{equation}
where $\Omega_{\rm ref}$ denotes the angular velocity at the rotation axis and $\varepsilon$ gives the solar \clem{($\varepsilon > 0$)} or anti-solar \clem{($\varepsilon < 0$)} behavior of the differential rotation.

\begin{figure}
\centering
\includegraphics[width=0.8\linewidth]{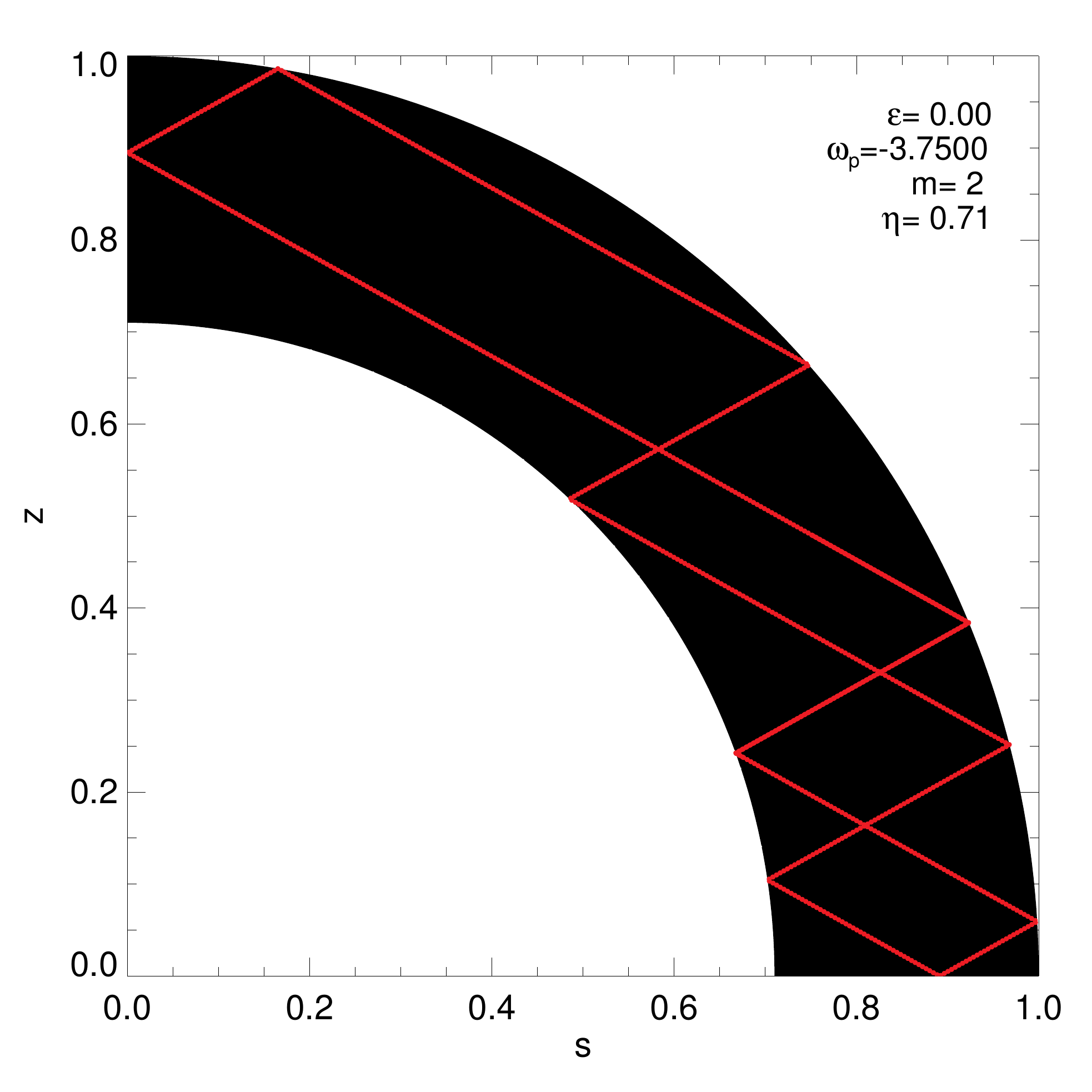} 
\includegraphics[width=0.8\linewidth]{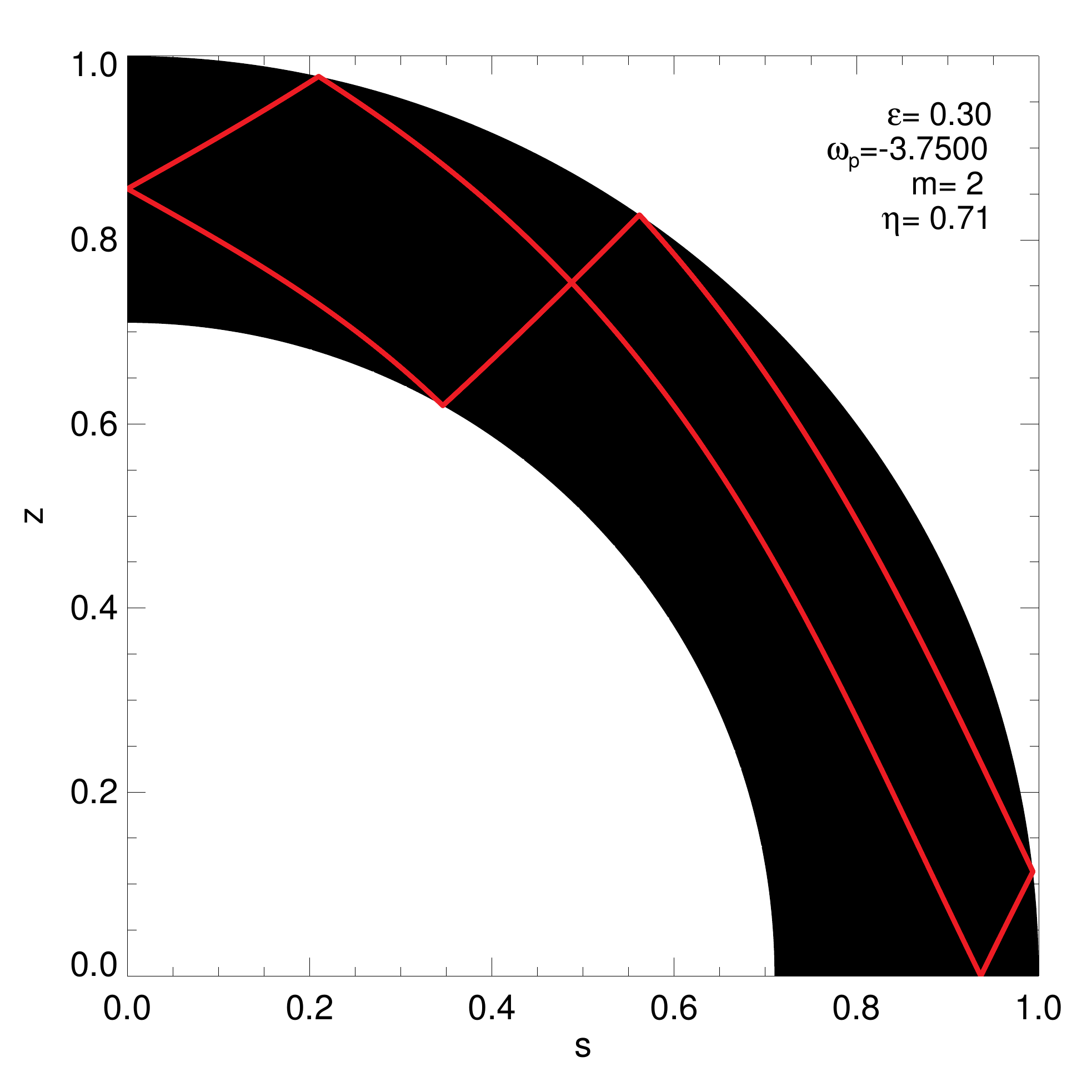}
\includegraphics[width=0.8\linewidth]{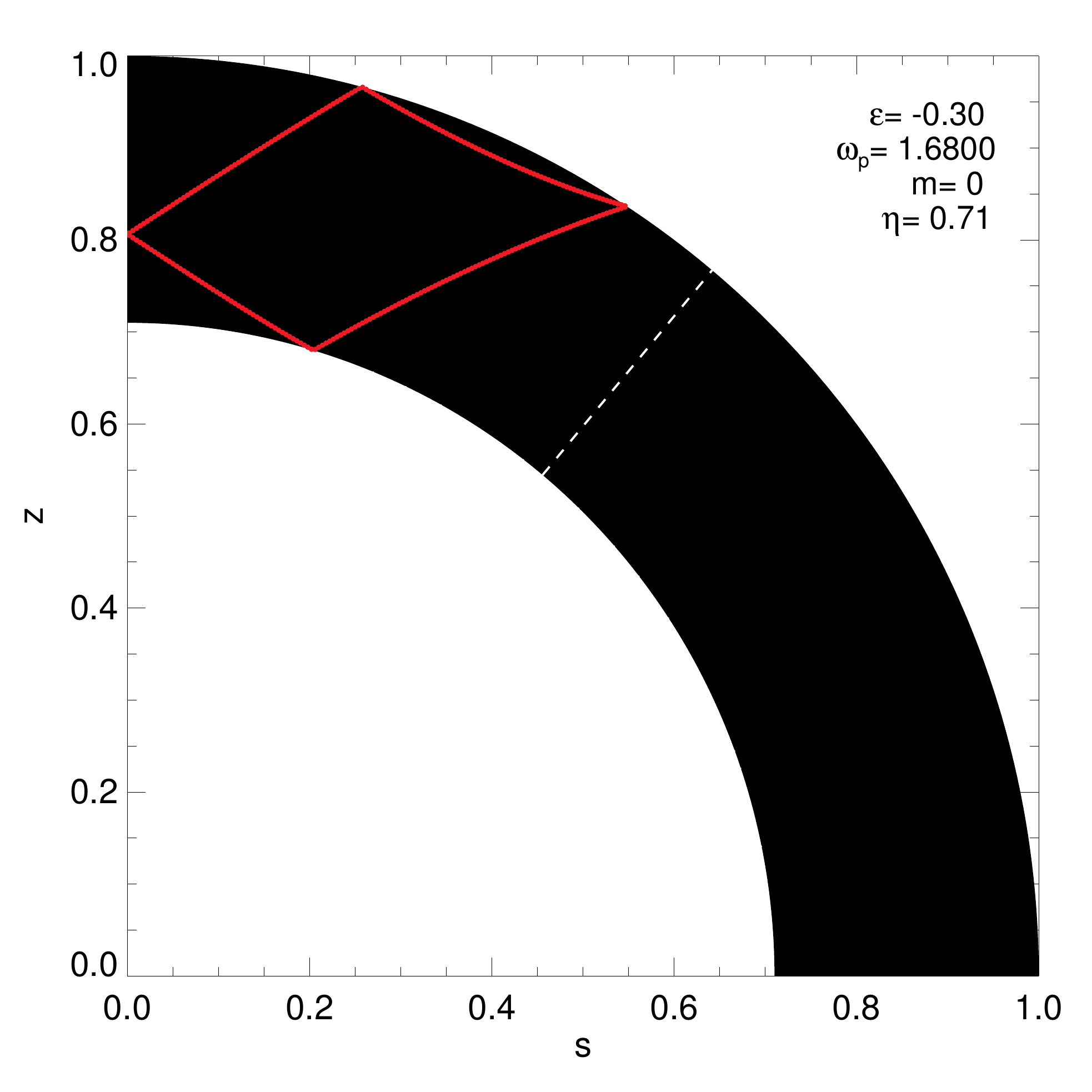}
\caption{\clem{Example of paths of characteristics for inertial eigenmodes in a meridional slice of a star's convective envelope {with a solar aspect ratio $\eta=0.71$}. {\bf Top~:} Attractor cycle \clem{with solid-body rotation} ($\varepsilon=0$). {\bf Center~:} Same for a D mode with solar-like conical rotation ($\varepsilon=0.3$). {\bf Bottom:} Same for a DT mode with anti-solar conical rotation ($\varepsilon =-0.1$), the dotted line depicts the turning surface. In all three plots, $m=2$}.}
\label{fig:attractors}
\end{figure}

\begin{figure}
\centering
\includegraphics[width=0.9\linewidth]{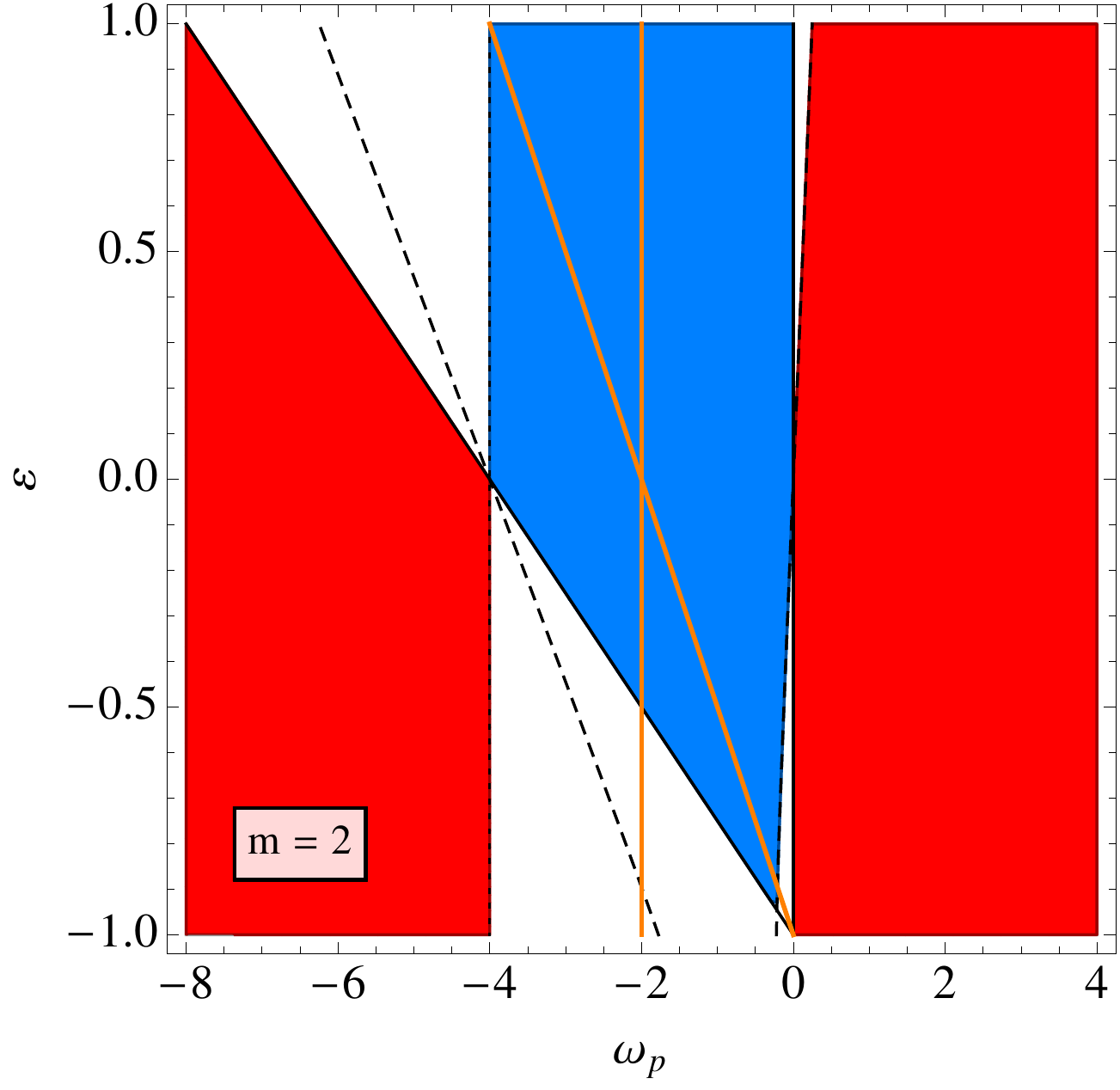}
\caption{Illustration of the two kinds of inertial modes propagating in \clem{a convective envelope} with conical rotation profile (see Eq. \ref{eq:rotation_profile}). {\bf Blue~:} D modes can propagate in the whole shell. {\bf White~:} DT modes exhibit at least one turning surface inside the shell. {\bf Red~:} No inertial modes \clem{can} propagate. {\bf Orange~:} The modes between the two orange lines feature a corotation \clem{resonance} $(\tilde{\omega}_{p} = 0)$ in the shell.}
	\label{fig:BBR}
\end{figure}

\clem{In the inviscid eigenvalue problem,} we can derive \clem{the propagation properties of free inertial waves upon examination at the paths of characteristics}. \clem{For conical differential rotation ($\varepsilon\neq 0$)}, we find that \citep{Guenel2016}:
\begin{itemize}
\item the slope of characteristics depends on position, which means their \clem{paths are curves} instead of straight lines \clem{as in solid body rotation} (see fig. \ref{fig:attractors}),
\item \clem{the usual range of Doppler-shifted frequencies allowed for inertial waves {in solid body rotation}, $\tilde\omega_{\rm p} = \omega_{\rm p} + m\Omega_{\rm ref} \in \left[-2\Omega_{\rm ref},2\Omega_{\rm ref}\right]$,} can be broadened by differential rotation 
(see fig. \ref{fig:BBR}),
\item \clem{turning surfaces arising because of differential rotation imply two kinds of eigenmodes: D modes (where inertial waves can propagate throughout the entire convective envelope) and DT modes (waves propagate in part of the envelope)},
\item corotation \clem{resonances} exist in the shell \clem{(they are locations where $\tilde\omega_{\rm p} = 0$)},
\item paths of characteristics may focus \clem{along attractor cycles or toward single points which are located} at the intersection of a turning surface \clem{and one of the shell's boundaries}.
\end{itemize}
The variety of inertial eigenmodes is \clem{illustrated} in fig. \ref{fig:BBR} for \clem{the azimuthal wavenumber} $m=2$. \clem{We have examined their dissipation properties in the viscous eigenvalue problem $(\nu \neq 0)$ by solving the viscous linear equations using a spectral code}. For more details, see \cite{Guenel2016}.

\section{Tidally-forced regime and viscous dissipation}
\label{sec:forced}

\subsection{Physical setup and simplified forcing}
\label{sec:forcing_model}
\clem{We now present our first results on tidally excited inertial oscillations in the convective envelope of a low-mass star with conical differential rotation. We look for perturbations of the velocity ($\bf u$) and of the reduced pressure ($p$) that are proportional to $\exp\left(i\omega_{p} t + im\varphi\right)$ and which} satisfy the linear system :
\begin{equation}
i \tilde{\omega}_p{\bf u} + 2 \Omega {\bf e_z} \times {\mathbf u} + r \sin \theta \left( {\bf u}\cdot\nabla\Omega \right) {\mathbf e_{\varphi}} = -{\nabla}p + \nu \Delta {\bf u} + {\bf f},
\label{eq:momentum}
\end{equation}
\begin{equation}
\nabla \cdot {\bf u} = 0,
\label{eq:continuity}
\end{equation}
along with stress-free boundary conditions. In these equations, $\bf f$ is the tidal forcing, namely a harmonic forcing with tidal frequency $\omega_p$ in the inertial frame ($\tilde{\omega}_p(\theta) = \omega_p+m\Omega(\theta)$ is the Doppler-shifted tidal frequency in the fluid frame).

\clem{In this work we adopt the same modelling of the tidal forcing as in \cite{Ogilvie2009}. Instead of including an explicit tidal force ($\bf f$) in Eq.~(\ref{eq:momentum}), we set the radial component of the perturbed radial velocity at the surface of the shell to scale as a single spherical harmonic (the scaling factor, which we denote by $A$, is arbitrary). We choose the harmonic $Y_2^2(\theta, \varphi)$, which corresponds to the case of a circular and coplanar orbit.} This assumption implies \clem{that we calculate} forced inertial oscillations with $m=2$ only.

\subsection{Energy balance and viscous dissipation}
Using the same spectral method as in \cite{Rieutord1987} and \cite{Guenel2016}, we compute numerical \clem{solutions} to Eqs. (\ref{eq:momentum})-(\ref{eq:continuity}) \clem{varying the shell's} aspect ratio $\eta$, the differential rotation \clem{parameter} $\varepsilon$, the Ekman number $\rm E = \nu / \left(\Omega_{ref} R^2\right)$ (\clem{with $R$ the shell's radius)} and the tidal frequency $\omega_{p}$.

For all these calculations, we compute the (time-averaged) volume-integrated viscous dissipation rate 
\begin{equation}
\left< D_{\nu} \right>_{T} = \frac{1}{2} D_{\nu} = \int_{V} \rho \, \nu \, \Delta {\bf u} \cdot {\bf u}^* \, {\rm d}V
\label{eq:viscous_dissipation}
\end{equation}
as a proxy for tidal dissipation. In the case of solid-body rotation, this quantity is equal to the (time-averaged) power input by the tidal forcing. But this does not hold true when differential rotation is present $(\varepsilon \neq 0)$ \clem{as free energy is also supplied from the shear}.

As a first step, we study the influence of viscosity (\clem{embodied in the} Ekman number $\rm E$) \clem{for} a Sun-like star with aspect ratio $\eta=0.71$ and $\varepsilon = 0.3$. For these parameters, the frequency range in which inertial eigenmodes may exist is \clem{$-4.6 \leq \omega_{p}/\Omega_{\rm ref} \lesssim 0.075$}, but \clem{that of} D modes is restricted to \clem{$-4 \leq \omega_{p}/\Omega_{\rm ref} \leq 0$} (see \clem{Fig.~}\ref{fig:BBR}). \clem{Note that} the frequency range where corotation resonances exist is \clem{$-2.6 \leq \omega_{p}/\Omega_{\rm ref}  \leq -2$}.

\begin{figure}
\centering
\includegraphics[width=0.9\linewidth]{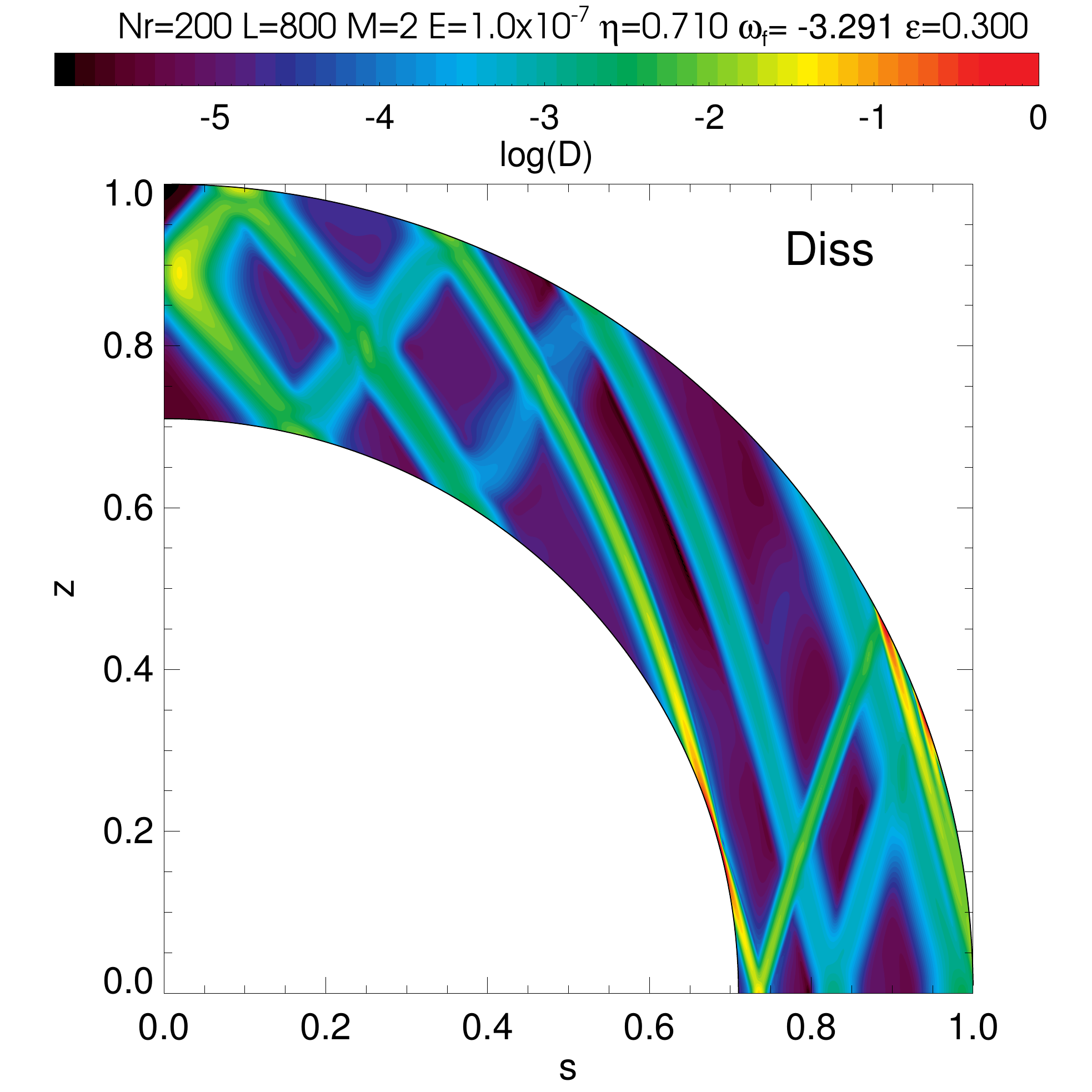}
\caption{Meridional cut \clem{of the viscous dissipation normalized to its maximum value for} a forced inertial oscillation with $m=2$, forcing frequency \clem{$\omega_{p} \approx -3.29\Omega_{\rm ref}$} and ${\rm E} = 10^{-7}$ in a Sun-like star $(\eta=0.71, \varepsilon=0.3)$.}
\label{fig:dissipation_modeD}
\end{figure}
\begin{figure*}
\centering
\includegraphics[width=\linewidth]{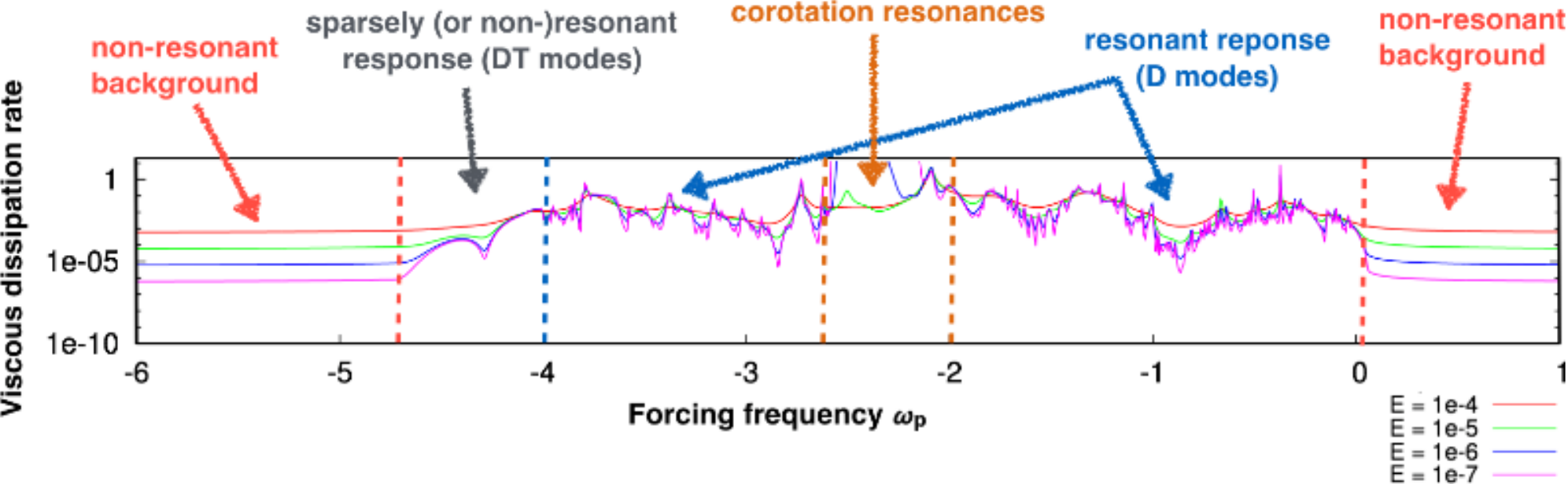}
\caption{Variation of the volume-integrated viscous dissipation \clem{rate} (in logarithmic scale in units of $\rho \, R^3 \, \Omega_{\rm ref} \, A^{2} $) as a function of tidal frequency $\omega_{p}$ \clem{(in units of $\Omega_{\rm ref}$)} in a Sun-like star $(\eta=0.71, \varepsilon=0.3)$ for different values of the Ekman number ${\rm E} = \{10^{-4}, 10^{-5}, 10^{-6}, 10^{-7}\}$.}
\label{fig:dissipation_spectrum}
\end{figure*}
For every Ekman number in $\{10^{-4}, 10^{-5}, 10^{-6}, 10^{-7}\}$ {relevant for stellar interiors \citep{OgilvieLin2007,Mathisetal2016}}, we \clem{calculate the fluid's response to the simplified forcing described in Sect.~\ref{sec:forcing_model}} for 1000 forcing frequencies in the range \clem{$-6 \leq \omega_{p}/\Omega_{\rm ref} \leq 1$}. For instance, Fig. \ref{fig:dissipation_modeD} displays a meridional cut of the viscous dissipation induced by a forced oscillation for $\rm E = 10^{-7}$ and \clem{$\omega_{p} \approx -3.29\Omega_{\rm ref}$} (in the \clem{frequency range of D modes}). \clem{From these calculations we infer} the (time-averaged) volume-integrated viscous dissipation rate defined by Eq. \ref{eq:viscous_dissipation}.  The viscous dissipation spectrum is displayed in Fig. \ref{fig:dissipation_spectrum}. First, we notice that the dissipation rate is almost constant outside of the range \clem{$-4.6 \leq \omega_{p}/\Omega_{\rm ref}  \lesssim 0.075$} \clem{since no inertial waves may propagate and thus dissipate in the shell}. This \clem{nearly} constant non-resonant background is usually the smallest value obtained for the dissipation rate over all tidal frequencies, and we checked that it scales with $\rm E$ \citep[e.g][]{OgilvieLin2004, AuclairDesrotour2015}. {In a real star, it may correspond to the so-called equilibrium/non-wave like tide \citep[e.g.,][]{Zahn1966,RMZ2012,Ogilvie2013}.}

In the frequency range \clem{$-4 \leq \omega_{p}/\Omega_{\rm ref}  \leq 0$} corresponding to D modes, \clem{the viscous dissipation spectrum} shows a behavior \clem{that is essentially reminiscent to the case of solid-body rotation}: it becomes \clem{richer} as $\rm E$ decreases, \clem{with an increasing number of peaks and troughs} \citep[e.g.,][]{Ogilvie2009, Rieutord2010,AuclairDesrotour2015}. Moreover, the \clem{viscous} dissipation rate seems to show very few \clem{resonant} peaks in the frequency range of DT modes (\clem{$-4.6 \leq \omega_{p}/\Omega_{\rm ref}  \leq -4$} and \clem{$0 \leq \omega_{p}/\Omega_{\rm ref}  \lesssim 0.075$}), which seems to confirm the analysis presented in \cite{Guenel2016}: \clem{with conical differential rotation}, there are few DT eigenmodes that can be excited by an external forcing, and the resulting viscous dissipation rate does not show any resonant behavior and remains small.

Finally, in the frequency range \clem{$-2.6 \leq \omega_{p}/\Omega_{\rm ref}  \leq -2$,} where a corotation resonance \clem{exists}, \clem{we find that the viscous dissipation rate is dramatically enhanced (by several orders of magnitude)} for \clem{${\rm E} = 10^{-6}$ and $10^{-7}$. However, no significant enhancement occurs for} ${\rm E}=10^{-4}$ and $10^{-5}$, despite the existence of the very same corotation resonance. \clem{We checked the convergence of our results against increasing the spectral resolution for the models where a corotation resonance threads the shell. While an enhanced viscous dissipation rate can be expected when a corotation resonance threads the shell (the flow is formally singular there in the inviscid limit, but the singularity is regularized by viscosity), the magnitude of this enhancement is not totally understood, in part due to the possibility that {some} inertial eigenmodes with conical differential rotation may be unstable in the presence of a corotation resonance \citep{Guenel2016}. Hence the cut in the y-axis in the dissipation spectrum displayed in Fig. \ref{fig:dissipation_spectrum}. Nonetheless, preliminary calculations with cylindrical differential rotation, where eigenmodes are always stable in the presence of a corotation resonance \citep{BR2013}, also show a large increase in the viscous dissipation rate in the frequency range of corotation resonances.{ This points towards a potential key role of corotation resonances for tidal dissipation in differentially rotating stars and planets.}}



\section{\clem{Concluding remarks}}
\clem{We have carried out a first numerical exploration of tidally excited inertial waves in the differentially rotating convective envelope of a low-mass star, assuming conical differential rotation {as observed in the Sun and other solar-like stars and predicted by numerical simulations}. Our preliminary results show a rich spectrum for the viscous dissipation rate of the waves. Interestingly, when a corotation resonance exists in the envelope, the viscous dissipation is greatly enhanced, which could lead a priori to a rapid tidal evolution of the orbit and/or spin of the system. Note however that in our study the background profile of differential rotation is fixed {\b and} not altered by the dissipation of tidal waves. Our work is to some extent  complementary to that of \cite{Favier2014},} who studied the deposition of angular momentum by forced inertial waves in the non-linear regime, but propagating in {an initially} \clem{uniformly-rotating} background flow. \clem{Future extensions of our work include taking into account the feedback of tidal waves' dissipation on the background rotation profile.}


\section*{Acknowledgments}
M. Guenel and S. Mathis acknowledge funding by the European Research Council through ERC grant SPIRE 647383. This work was also supported by the Programme National de Plan\'etologie (CNRS/INSU) and CNES PLATO grant at CEA-Saclay.

\bibliographystyle{cs19proc}
\bibliography{GMBR2016_CS19}

\begin{thebibliography}{25}
\providecommand{\natexlab}[1]{#1}

\bibitem[\protect\astroncite{{Auclair-Desrotour}
  \emph{et~al.}}{2014}]{ADLPM2014}
{Auclair-Desrotour}, P., {Le Poncin-Lafitte}, C., \& {Mathis}, S. 2014, \aap,
  561, L7.

\bibitem[\protect\astroncite{{Auclair Desrotour}
  \emph{et~al.}}{2015}]{AuclairDesrotour2015}
{Auclair Desrotour}, P., {Mathis}, S., \& {Le Poncin-Lafitte}, C. 2015, \aap,
  581, A118.

\bibitem[\protect\astroncite{{Barker} \& {Ogilvie}}{2009}]{BO2009}
{Barker}, A.~J. \& {Ogilvie}, G.~I. 2009, \mnras, 395, 2268.

\bibitem[\protect\astroncite{{Barnes} \emph{et~al.}}{2005}]{Barnesetal2005}
{Barnes}, J.~R., {Collier Cameron}, A., {Donati}, J.-F., {James}, D.~J.,
  {Marsden}, S.~C., \emph{et~al.} 2005, \mnras, 357, L1.

\bibitem[\protect\astroncite{{Baruteau} \& {Rieutord}}{2013}]{BR2013}
{Baruteau}, C. \& {Rieutord}, M. 2013, Journal of Fluid Mechanics, 719, 47.

\bibitem[\protect\astroncite{{Bolmont} \& {Mathis}}{2016}]{BM2016}
{Bolmont}, E. \& {Mathis}, S. 2016, Celestial Mechanics and Dynamical
  Astronomy, 126, 275.

\bibitem[\protect\astroncite{{Brun} \emph{et~al.}}{2015}]{Brun2015}
{Brun}, A.~S., {Garc{\'{\i}}a}, R.~A., {Houdek}, G., {Nandy}, D., \&
  {Pinsonneault}, M. 2015, \ssr, 196, 303.

\bibitem[\protect\astroncite{{Favier} \emph{et~al.}}{2014}]{Favier2014}
{Favier}, B., {Barker}, A.~J., {Baruteau}, C., \& {Ogilvie}, G.~I. 2014,
  \mnras, 439, 845.

\bibitem[\protect\astroncite{{Gastine} \emph{et~al.}}{2014}]{Gastine2014}
{Gastine}, T., {Yadav}, R.~K., {Morin}, J., {Reiners}, A., \& {Wicht}, J. 2014,
  \mnras, 438, L76.

\bibitem[\protect\astroncite{{Guenel} \emph{et~al.}}{2016}]{Guenel2016}
{Guenel}, M., {Baruteau}, C., {Mathis}, S., \& {Rieutord}, M. 2016, \aap, 589,
  A22.

\bibitem[\protect\astroncite{{Hut}}{1980}]{Hut1980}
{Hut}, P. 1980, Astronomy \& Astrophysics, 92, 167.

\bibitem[\protect\astroncite{{Mathis}}{2015}]{Mathis2015}
{Mathis}, S. 2015, \aap, 580, L3.

\bibitem[\protect\astroncite{{Mathis} \emph{et~al.}}{2016}]{Mathisetal2016}
{Mathis}, S., {Auclair-Desrotour}, P., {Guenel}, M., {Gallet}, F., \& {Le
  Poncin-Lafitte}, C. 2016, \aap, 592, A33.

\bibitem[\protect\astroncite{{Matt} \emph{et~al.}}{2011}]{Matt2011}
{Matt}, S.~P., {Do Cao}, O., {Brown}, B.~P., \& {Brun}, A.~S. 2011,
  Astronomische Nachrichten, 332, 897.

\bibitem[\protect\astroncite{{Meibom} \& {Mathieu}}{2005}]{Meibom2005}
{Meibom}, S. \& {Mathieu}, R.~D. 2005, \apj, 620, 970.

\bibitem[\protect\astroncite{{Ogilvie}}{2009}]{Ogilvie2009}
{Ogilvie}, G.~I. 2009, \mnras, 396, 794.

\bibitem[\protect\astroncite{{Ogilvie}}{2013}]{Ogilvie2013}
{Ogilvie}, G.~I. 2013, \mnras, 429, 613.

\bibitem[\protect\astroncite{{Ogilvie} \& {Lin}}{2004}]{OgilvieLin2004}
{Ogilvie}, G.~I. \& {Lin}, D.~N.~C. 2004, \apj, 610, 477.

\bibitem[\protect\astroncite{{Ogilvie} \& {Lin}}{2007}]{OgilvieLin2007}
{Ogilvie}, G.~I. \& {Lin}, D.~N.~C. 2007, \apj, 661, 1180.

\bibitem[\protect\astroncite{{Remus} \emph{et~al.}}{2012}]{RMZ2012}
{Remus}, F., {Mathis}, S., \& {Zahn}, J.-P. 2012, \aap, 544, A132.

\bibitem[\protect\astroncite{{Rieutord}}{1987}]{Rieutord1987}
{Rieutord}, M. 1987, Geophysical and Astrophysical Fluid Dynamics, 39, 163.

\bibitem[\protect\astroncite{{Rieutord} \& {Valdettaro}}{2010}]{Rieutord2010}
{Rieutord}, M. \& {Valdettaro}, L. 2010, Journal of Fluid Mechanics, 643, 363.

\bibitem[\protect\astroncite{{Schou} \emph{et~al.}}{1998}]{Schouetal1998}
{Schou}, J., {Antia}, H.~M., {Basu}, S., {Bogart}, R.~S., {Bush}, R.~I.,
  \emph{et~al.} 1998, \apj, 505, 390.

\bibitem[\protect\astroncite{{Varela} \emph{et~al.}}{2016}]{Varelaetal2016}
{Varela}, J., {Strugarek}, A., \& {Brun}, A.~S. 2016, Advances in Space
  Research, 58, 1507.

\bibitem[\protect\astroncite{{Zahn}}{1966}]{Zahn1966}
{Zahn}, J.~P. 1966, Annales d'Astrophysique, 29, 313.

\end{thebibliography}

\end{document}